\documentclass{article}
\usepackage{spconf,amsmath,graphicx,color}

\usepackage{enumitem}
\setlist{nosep, leftmargin=14pt}

\usepackage{mwe} 

\title{SS-CADA: A S\MakeLowercase{emi}-s\MakeLowercase{upervised} c\MakeLowercase{ross}-a\MakeLowercase{natomy} d\MakeLowercase{omain} a\MakeLowercase{daptation} \MakeLowercase{for} c\MakeLowercase{oronary} a\MakeLowercase{rtery} s\MakeLowercase{egmentation}}
\name{Jingyang Zhang$^{1,4,\star}$, \quad Ran Gu$^{2,\star}$, \quad Guotai Wang$^{2,\dagger}$, \quad Hongzhi Xie$^{3}$, \quad Lixu Gu$^{1,4,\dagger}$ 
\thanks{$\star$ The authors contributed equally. $\dagger$ The corresponding authors: Guotai Wang (guotai.wang@uestc.edu.cn) and Lixu Gu (gulixu@sjtu.edu.cn).}}

\address{$^{1}$ School of Biomedical Engineering, Shanghai Jiao Tong University, Shanghai, China\\
    $^{2}$ School of Mechanical and Electrical Engineering, University of Electronic Science and \\ Technology of China, Chengdu, China\\
    $^{3}$ Department of Cardiology, Peking Union Medical College Hospital, Beijing, China\\
    $^{4}$ Institute of Medical Robotics, Shanghai Jiao Tong University, Shanghai, China}

\begin{document}
%
\maketitle
\begin{abstract}
The segmentation of coronary arteries by convolutional neural network is promising yet requires a large amount of labor-intensive manual annotations.
Transferring knowledge from retinal vessels in widely-available public labeled fundus images (FIs) has a potential to reduce the annotation requirement for coronary artery segmentation in X-ray angiograms (XAs) due to their common tubular structures.
However, it is challenged by the \emph{cross-anatomy domain shift} due to the intrinsically different vesselness characteristics in different anatomical regions under even different imaging protocols.
To solve this problem, we propose a \underline{S}emi-\underline{S}upervised \underline{C}ross-\underline{A}natomy \underline{D}omain \underline{A}daptation (SS-CADA) which requires only limited annotations for coronary arteries in XAs.
With the supervision from a small number of labeled XAs and publicly available labeled FIs, we propose a vesselness-specific batch normalization (VSBN) to individually normalize feature maps for them considering their different cross-anatomic vesselness characteristics. In addition, to further facilitate the annotation efficiency, we employ a self-ensembling mean-teacher (SE-MT) to exploit abundant unlabeled XAs by imposing a prediction consistency constraint.
Extensive experiments show that our SS-CADA is able to solve the challenging cross-anatomy domain shift, achieving accurate segmentation for coronary arteries given only a small number of labeled XAs.
\end{abstract}
\section{Introduction}
\label{sec:intro}

The segmentation of coronary arteries in X-ray angiograms (XAs) is indispensable for both coronary artery disease diagnosis and percutaneous coronary intervention navigation. 
Convolutional neural networks have achieved remarkable progress in this task \cite{AR-SPL,seg1}, yet requiring a large amount of manual annotations for training images, which is highly time-consuming and labor-intensive due to the complex coronary-tree structures with varying lumen sizes and inhomogeneous contrast agent inflow. Therefore, it is desired to reduce the manual annotations for coronary arteries in XAs while maintaining the segmentation performance.

\begin{figure}[t]
    \vspace{-0.2cm}
    \setlength{\abovecaptionskip}{-0cm}
    \setlength{\belowcaptionskip}{-0.2cm}
    \centerline{\includegraphics[width=\columnwidth]{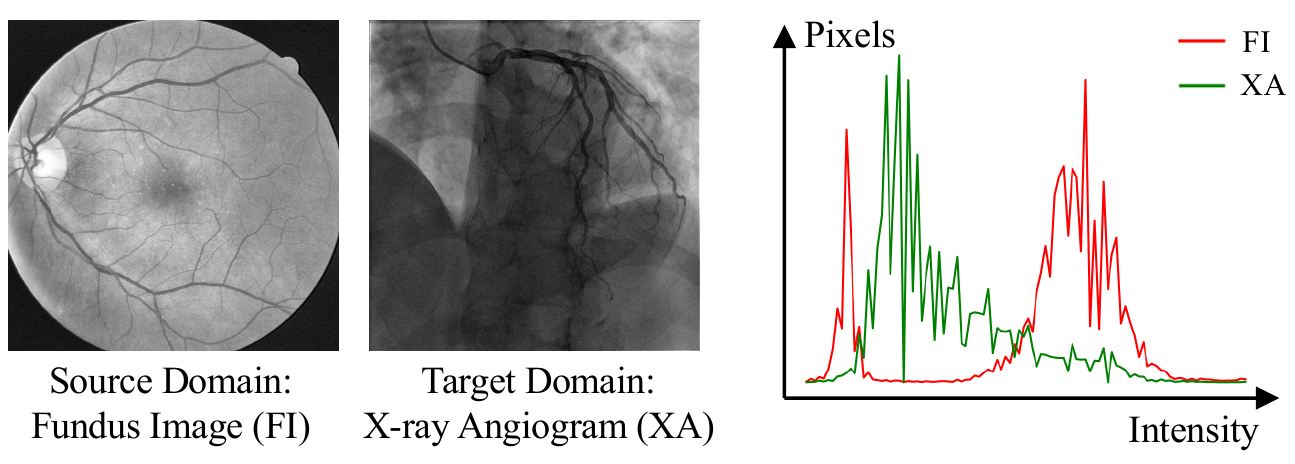}}
    \caption{Illustration of the challenging cross-anatomy domain shift between fundus image (FI) and X-ray angiogram (XA).}
    \label{domain_shift}
\end{figure}

Considering that the tubular feature of coronary artery is similar to that of vascular structures in other part of the body, e.g., retinal vessels in fundus images (FIs), and there are several publicly available annotated FI datasets, e.g., DRIVE~\cite{DRIVE}, STARE~\cite{STARE} and RITE~\cite{RITE}, it is promising to transfer the knowledge from annotated FIs to facilitate the coronary artery segmentation task in XAs. These publicly available annotated FI data is regarded as a well-established source domain, which enables a potentially cost-free knowledge transfer to the target domain (i.e., XAs).
However, data distributions of FIs and XAs mismatch significantly due to their intrinsically different vesselness characteristics in different anatomical regions (i.e., the retinal fundus and the heart) under different imaging protocols (i.e., optical and radiological imaging), which is called the \emph{cross-anatomy domain shift} as shown in Fig. \ref{domain_shift}. 
On the contrary, existing unsupervised domain adaptation (UDA) methods \cite{MMD,UDA_O,Pnp_adanet,SIFA} are devoted to solving only cross-modality domain shift, i.e., the same anatomical region under different imaging modalities. They align the feature space (or even the image appearance \cite{SIFA}) across modalities explicitly by minimizing the maximum mean discrepancy \cite{MMD} and implicitly by adversarial learning \cite{UDA_O,Pnp_adanet}, without using target domain labels. 
However, compared with such common cross-modality domain shift, the cross-anatomy discrepancy in our task is much more challenging and under-studied. Yu \emph{et al.} \cite{SC_UDA} presented the first attempt to mitigate the cross-anatomy discrepancy by forcing shape consistency in UDA yet limited by the lack of reliable synthetic labels for semantic guidance. Therefore, existing UDA methods tend to be misguided by the intractable cross-anatomy domain shift between FIs and XAs, leading to performance degradation for coronary artery segmentation.

To solve this problem, we propose a \underline{S}emi-\underline{S}upervised \underline{C}ross-\underline{A}natomy \underline{D}omain \underline{A}daptation (SS-CADA) for coronary artery segmentation. Only a few labeled XAs are required, while the public labeled FIs and abundant unlabeled XAs can be obtained without annotation cost. Considering the cross-anatomical vesselness difference of coronary arteries and retinal vessels, we propose a vesselness-specific batch normalization (VSBN) to individually normalize the feature maps of them by using the supervision from a small number of labeled XAs and publicly avaliable FIs. In addition, to further facilitate the annotation efficiency, we adopt a self-ensembling mean-teacher (SE-MT) architecture to leverage abundant unlabeled XAs by encouraging the prediction consistency with random perturbations. 
The proposed SS-CADA is compact and flexible without using adversarial learning \cite{SSDA} that suffers from the intractable Nash equilibrium solution \cite{nash_point}, the additional network designs and the complex training process.
Extensive experiments demonstrate that our SS-CADA achieves accurate segmentation for coronary arteries given only very limited manual annotations for XAs.

\section{Methodology}
\label{sec:method}

\begin{figure}[t]
    \centering
    \vspace{-0.2cm}
    \setlength{\abovecaptionskip}{-0cm}
    \setlength{\belowcaptionskip}{-0.2cm}
    \includegraphics[width=0.5\textwidth]{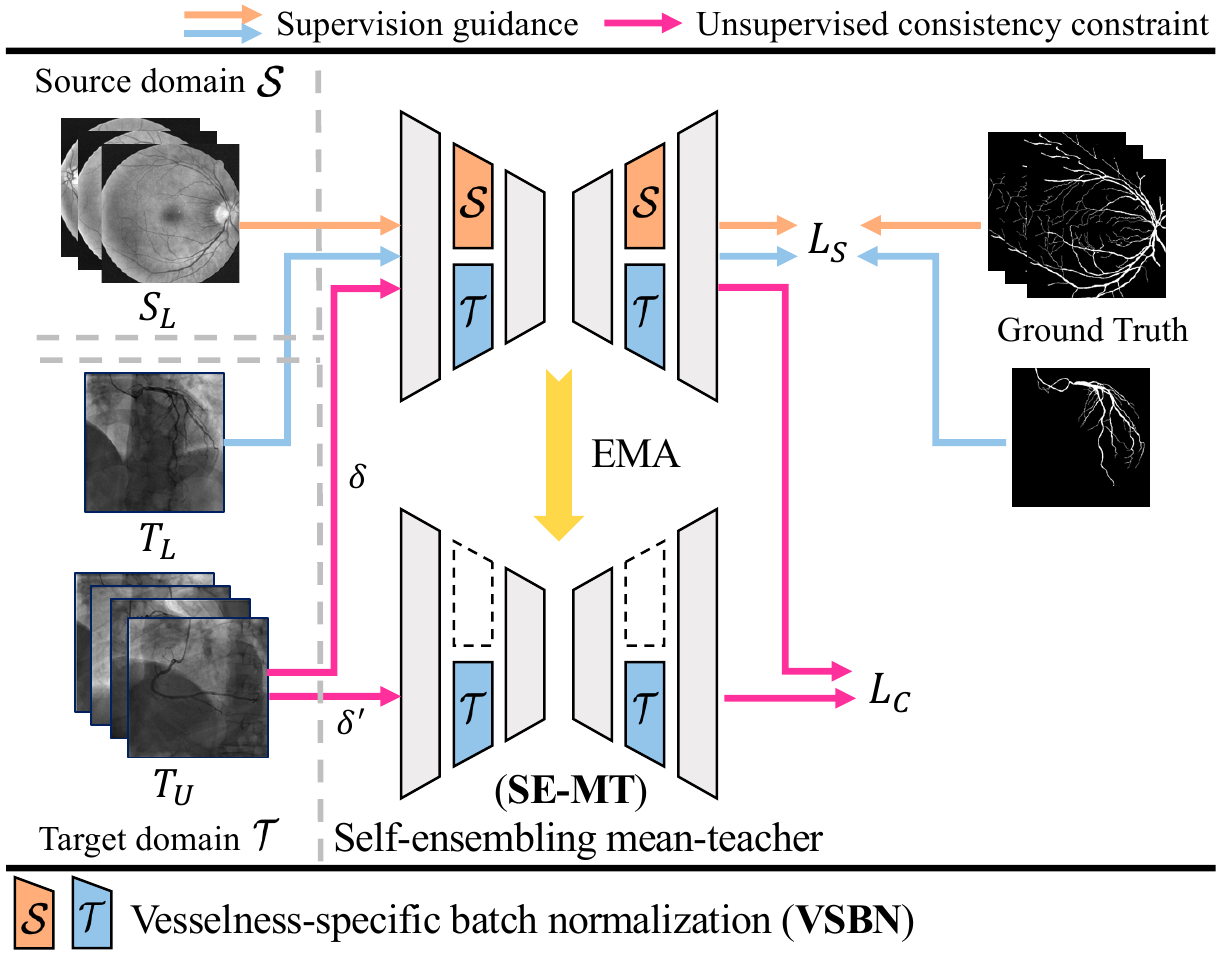}
    \caption{Flow chart of the proposed SS-CADA that consists of two parts: 1) a segmentation network with vesselness-specific batch normalization (VSBN); and 2) a self-ensembling mean-teacher (SE-MT) architecture.}
    \label{fig:model}
\end{figure}

Let $\mathcal{S}$ and $\mathcal{T}$ denote the source domain of FIs with retinal vessels and the target domain of XAs with coronary arteries, respectively.
We are given publicly available annotated FIs $S_L=\{(x_{i}^{S_L}, y_{i}^{S_L})\}^{n^{S_L}}_{i=1}\}$, a small set of annotated XAs $T_L=\{(x_{i}^{T_L}, y_{i}^{T_L})\}^{n^{T_L}}_{i=1}\}$ and abundant unlabeled XAs $T_U=\{(x_{i}^{T_U})\}^{n^{T_U}}_{i=1}$.
The proposed semi-supervised cross-anatomy domain adaptation (SS-CADA) is depicted in Fig.~\ref{fig:model}, which consists of two parts: 1) a segmentation network with vesselness-specific batch normalization (VSBN) that learns from $S_L$ and $T_L$ to provide a supervision guidance to bridge the cross-anatomy discrepancy; and 2) a self-ensembling mean-teacher (SE-MT) that imposes an unsupervised consistency constraint on $T_{U}$ for more annotation efficiency.

\subsection{Vesselness-specific batch normalization (VSBN)}\label{sec:VSBN}
\begin{figure}[t]
    \centering
    \vspace{-0.2cm}
    \setlength{\abovecaptionskip}{-0cm}
    \setlength{\belowcaptionskip}{-0.2cm}
    \includegraphics[width=0.48\textwidth]{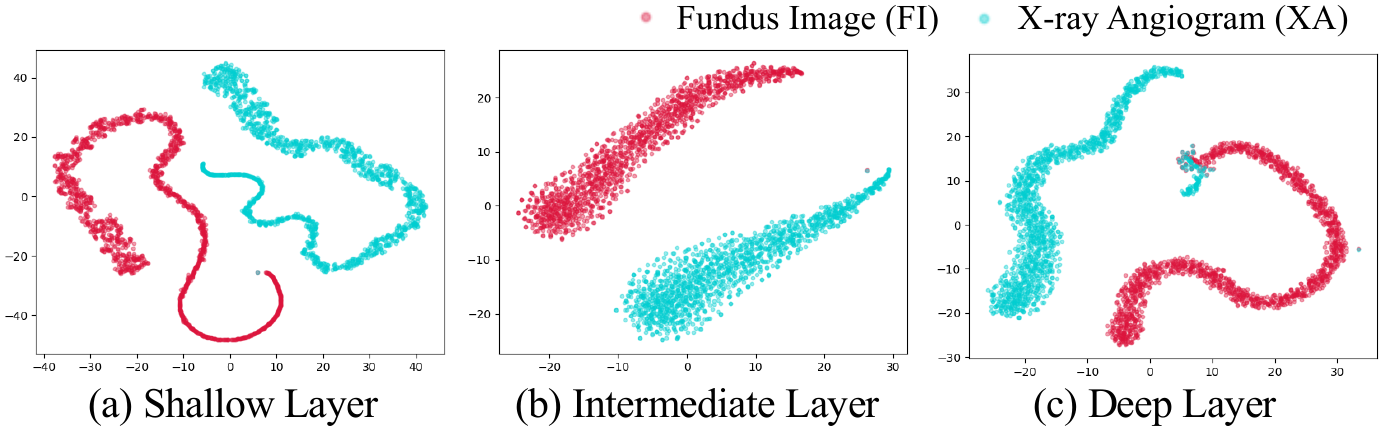}
    \caption{t-SNE \cite{TSNE} visualization of BN statistic distributions in shallow, intermediate and deep layer from FIs and XAs.}
    \label{fig:BN_tsne}
\end{figure}

Without loss of generality, we adopt a classical U-Net~\cite{unet} as the network backbone for segmentation. Batch normalization (BN)~\cite{BN} is used in the network to reduce internal convariate shift, leading to improved model generalization and faster convergence. 
In our cross-anatomy problem setting, it is interesting to observe that when two separated networks with BN layers are used to learn individually from $S_L$ and $T_L$, the BN statistic (batch-wise mean and variance) are highly separable, as shown in Fig.~\ref{fig:BN_tsne}. 
It suggests that the separable distribution of BN statistic is able to account for the cross-anatomy discrepancy between domain $\mathcal{S}$ and $\mathcal{T}$ to some extent. As a result, a joint training manner that directly mixes $S_L$ and $T_L$ would have limited performance, since the shared kernels are misguided by statistical variations between domain $\mathcal{S}$ and $\mathcal{T}$ and thus fail to grasp the generic representations.

To solve this problem, we propose a vesselness-specific batch normalization (VSBN) layer in the network, which adopts individual BN parameters for domain $\mathcal{S}$ and $\mathcal{T}$ considering their cross-anatomical vesselness difference. Meanwhile, convolutional kernels are shared for domain $\mathcal{S}$ and $\mathcal{T}$ to learn a general representation, such as the similar tubular feature. Formally, let $f^{d}=[f^{d}_1,...,f^{d}_{c},...,f^{d}_{C}]$ denote one feature map given an input from domain $d\in{\{\mathcal{S},\mathcal{T}\}}$, where $f^{d}_{c}$ is the feature map in channel $c$. 
The proposed VSBN normalizes each channel respectively and then applies affine transformation with trainable parameters that are specific to a certain domain $d$, i.e., rescale parameters $\gamma^{d}=[\gamma^{d}_1,...,\gamma^{d}_{c},...,\gamma^{d}_{C}]$ and bias parameters $\beta^{d}=[\beta^{d}_1,...,\beta^{d}_{c},...,\beta^{d}_{C}]$:
\begin{equation}
    \hat{f}_c^{d}=\gamma_c^{d}\cdot \overline{f}_c^{d} + \beta^{d}_{c},\quad \text { where } \quad \overline{f}_c^{d}=\frac{f_c^{d}-\mu_c^{d}}{\sqrt{(\sigma_c^{d})^2+\varepsilon}},
\end{equation}
where $\hat{f}_c^{d}$ is the channel $c$ of VSBN output. $\mu_c^{d}$ and $\sigma_c^{d}$ are the mean and standard deviation of $f_c^{d}$, and $\varepsilon$ is an infinitesimal.

The proposed VSBN enables a compact dual-model architecture to deal with FIs and XAs respectively using different parameter sets, i.e., $\Theta^{\mathcal{S}}=[\theta,\gamma^{\mathcal{S}},\beta^{\mathcal{S}}]$ for FIs and $\Theta^{\mathcal{T}}=[\theta,\gamma^{\mathcal{T}},\beta^{\mathcal{T}}]$ for XAs, which share the convolution parameters $\theta$ while use different BN parameters specific to FIs and XAs, respectively.
Given image-annotation pairs $(x_i^{S_L},y_i^{S_L})\in S_L$ from domain $\mathcal{S}$ and $(x_i^{T_L},y_i^{T_L})\in T_L$ from domain $\mathcal{T}$, we define a loss function to jointly optimize these parameter sets:
\begin{equation}
\label{EQ:loss_sup}
    L_{s} = \sum_{i=1}^{n^{S_L}}L_{seg}(p_i^{S_L},y_i^{S_L}) + \sum_{i=1}^{n^{T_L}}L_{seg}(p_i^{T_L},y_i^{T_L}),
\end{equation}
where $p_i^{S_L}=\psi(x_i^{S_L};\Theta^{\mathcal{S}})$ and $p_i^{T_L}=\psi(x_i^{T_L};\Theta^{\mathcal{T}})$ are predictions of $x_i^{S_L}$ and $x_i^{T_L}$ using the corresponding parameter sets $\Theta^{\mathcal{S}}$ and $\Theta^{\mathcal{T}}$. Moreover, $L_{seg}$ denotes a hybrid segmentation loss that consists of the cross-entropy loss and dice loss.

\subsection{Self-ensembling mean teacher (SE-MT)}
The proposed VSBN relieves the cross-anatomy domain shift by the supervision guidance of $S_L$ and $T_L$. However, the small number of annotated XAs in $S_L$ may limit the performance of the model.
To deal with this problem, we employ a self-ensembling mean teacher (SE-MT) architecture to exploit abundant unlabeled XAs in $T_U$ without annotation cost.
Specifically, we define the sub-model with parameter set $\Theta^{\mathcal{T}}$ specific to XAs as the student model. Then, we calculate an exponential moving average (EMA) for it and generate a teacher model, which produces better predictions due to the temporal self-ensembling \cite{temporal}. Finally, we define an unsupervised consistency loss between the predictions of student model and teacher model given the same input $x_i^{T_U}\in T_U$ from domain $\mathcal{T}$ yet under different perturbations $\delta$ and $\delta'$:
\begin{equation}
\label{Eq:loss_consistency}
    L_{c} = \sum_{i=1}^{n^{T_U}}L_{mse}\left( \psi(x_i^{T_U};\Theta^{\mathcal{T}},\delta), \psi(x_i^{T_U};(\Theta^{\mathcal{T}})',\delta')\right)
\end{equation}
where $(\Theta^{\mathcal{T}})'$ is the EMA result of $\Theta^{\mathcal{T}}$, and it is used for the teacher model. $\psi(x_i^{T_U};\Theta^{\mathcal{T}},\delta)$ and $\psi(x_i^{T_U};(\Theta^{\mathcal{T}})',\delta')$ are the student and teacher model prediction, respectively, which are used to compute the mean squared error loss $L_{mse}$.

Overall, integrating the supervised loss $L_s$ in Eq.~\ref{EQ:loss_sup} and the unsupervised consistency loss $L_c$ in Eq.~\ref{Eq:loss_consistency}, the complete training objective of the proposed SS-CADA is formulated as:
\begin{equation}
\label{Eq；objective}
    L = L_s + \lambda\cdot L_c
\end{equation}
where $\lambda$ acts as an ramp-up trade-off parameter. Once the training process is completed, an inference process obtains coronary artery segmentation result by performing forward propagation on the student model with parameter set $\Theta^{\mathcal{T}}$.

\begin{figure*}[t]
    \centerline{\includegraphics[width=7in]{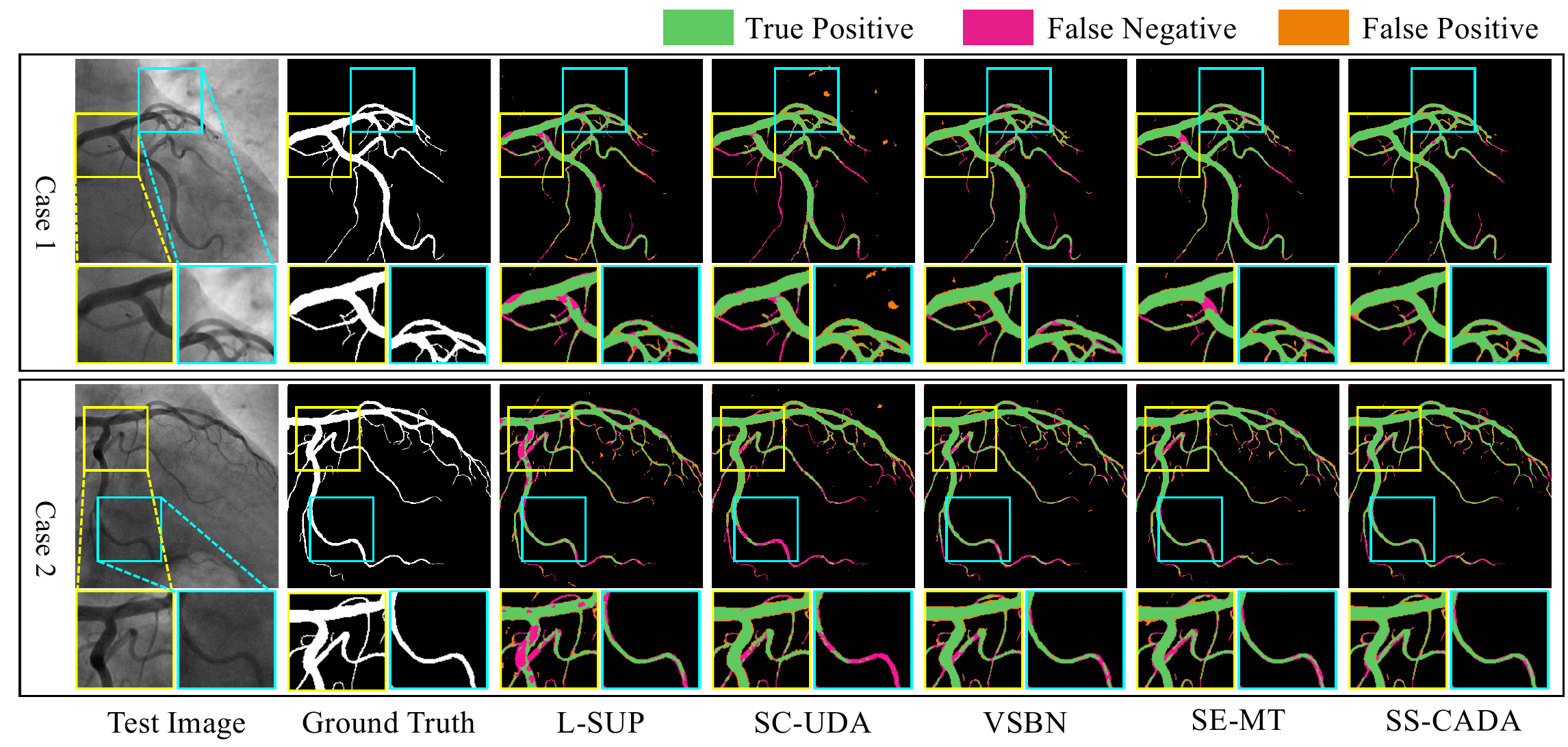}}
    \caption{Two examples of segmentation results. The true positives, false negatives and false positives are colored in green, red and orange, respectively. The zoomed views are appended below each case to highlight the segmentation details.}
    \label{segmentations}
\end{figure*}

\section{experiments and results}
\label{sec:experiments and results}
\subsection{Dataset and implementation details}
We used the FI dataset DRIVE \cite{DRIVE} with annotated retinal vessels as $S_L$ in source domain. It contains 40 FIs that were acquired from a Canon CR5 nonmydriatic 3CCD camera at 45$^o$ field of view. We used their green channel and pre-processed them by contrast limited adaptive histogram equalization and gamma correction.
In addition, we collected 191 XAs of 30 patients with coronary arteries as the target domain, using a Philips UNIQ FD10 C-arm system. Among them, for training, an expert radiologist randomly annotated 20 XAs of 3 patients that is regarded as $T_L$, and we took 92 XAs of 14 patients without annotations as $T_U$. The annotation process was performed on a PyQT GUI, where 5$\times$ resolution was used for the clear visualization of even thin branches. In addition, we used the remaining 43 XAs of 7 patients for validation and 36 XAs of 6 patients for testing. All FIs and XAs were resized to 512$\times$512 before feeding them in the network.

In SE-MT, the EMA decay rate was empirically set to 0.99 \cite{temporal}, and a time-dependent Gaussian warming up function $\lambda(t)=0.1 * e^{\left(-5\left(1-t / t_{max}\right)^{2}\right)}$ was used to dynamically change hyperparameter $\lambda$, where $t$ denotes the current training epoch and $t_{max}=50$ is the last epoch. Eq.~\ref{Eq；objective} was optimized with stochastic gradient descent with momentum 0.9, batch size 6, max iteration 30000 and initial learning rate 0.001 that was decayed exponentially with power 0.95.

\subsection{Results and analysis}
To demonstrate the effectiveness of our SS-CADA for solving cross-anatomy domain shift, we compare it with several other methods which can be divided into four categories: 1) using only $T_L$: A standard U-Net \cite{unet} is learned only from the small set of labeled XAs, which is denoted as L-SUP; 2) using $S_L$ and $T_U$: this is a standard UDA problem, and we use SC-UDA~\cite{SC_UDA} that incorporates shape consistency between two domains for this purpose; 3) using $S_L$ and $T_L$: this is a multi-domain learning problem, and we adopt three methods for this purpose such as a joint training paradigm (JOINT) without considering the domain discrepancy, a X-shape multi-modality learning network (MML)~\cite{xshape} and an U-Net with the proposed VSBN; and 4) using $T_L$ and $T_U$: this is a standard semi-supervised learning problem, which is addressed by the self-ensembling mean-teacher and denoted as SE-MT.

As shown in Table \ref{table}, the low Dice of L-SUP indicates that using only the small set of labeled XAs cannot obtain a sufficient model generalization capability. SC-UDA leads to even lower Dice when additional abundant unlabeled XAs are used, due to the misguidance of the challenging cross-anatomy domain shift. JOINT, MML and VSBN achieve better results than SC-UDA, demonstrating that a small set of labeled XAs can provide effective supervision guidance for bridging the cross-anatomy domain shift. However, they are still largely inferior to our SS-CADA due to the ignorance of unlabeled XAs. On the contrary, SE-MT uses only the knowledge specific to coronary arteries in XAs yet without transferable knowledge from retinal vessels in FIs, leading to 2.06$\%$ decrease in Dice compared with our SS-CADA. 
The comparison between VSBN, SE-MT and SS-CADA is also regarded as the ablation study, which validates the feasibility of each key component of our method. The highest recall, precision and Dice of our SS-CADA emphasize that the cross-anatomy domain shift can be effectively solved in an annotation-efficient manner, i.e., manual annotations are required for only 18\% of the XAs.

\begin{table}[t]
    \caption{Quantitative comparison with other methods, where the ablation study is also performed between VSBN, SE-MT and SS-CADA. The best results are highlighted in bold font.}
    \vspace{1mm}
    \scriptsize
    \centering
    \scalebox{1.05}{\begin{tabular}{lllll}
    \hline
    \hline
    Training Set & Method & Recall ({\%}) & Precision ({\%}) & Dice ({\%})\\
    \hline
    $T_L$ & L-SUP & 75.12$\pm$4.83 & 65.87$\pm$8.47 & 69.81$\pm$5.35\\
    \hline
    $S_L$,$T_U$ & SC-UDA & 80.87$\pm$7.69 & 57.88$\pm$8.73 & 66.92$\pm$6.83\\
    \hline
    & JOINT & 81.14$\pm$6.75 & 62.66$\pm$9.35 & 70.37$\pm$7.37\\
    $S_L$,$T_L$ & MML & 83.10$\pm$4.11 & 65.11$\pm$10.92 & 72.49$\pm$7.69\\
    &VSBN & 80.66$\pm$4.58 & 69.12$\pm$9.53 & 74.12$\pm$6.62\\
    \hline
    $T_L$,$T_U$ & SE-MT & 82.55$\pm$3.89 & 72.14$\pm$8.40 & 76.70$\pm$5.43\\
    \hline
    $S_L$,$T_L$,$T_U$ & {SS-CADA} & \textbf{83.27$\pm$3.95} & \textbf{75.12$\pm$6.59} & \textbf{78.84$\pm$4.60}\\
    \hline
    \hline
    \end{tabular}}
    \label{table}
\end{table}

We visualize the segmentation results in Fig. \ref{segmentations}. Due to the page limitation, among all methods that use $S_L$ and $T_L$, only VSBN is shown due to its highest Dice. SC-UDA exhibits obvious false negatives for terminal vessels with thin scale and attenuated contrast, as well as false positives scattered in semi-transparent background. 
Less false positives can be achieved in L-SUP, VSBN and SE-MT, while false negatives for bifurcation points and thin vessels seem to be intractable, breaking the normal connectivity in coronary-tree structures. Given the small set of labeled XAs, only the proposed SS-CADA enables accurate extraction of coronary arteries with the best connectivity and the least false positives.

\section{Conclusion}
This paper propose a semi-supervised cross-anatomy domain adaptation for coronary artery segmentation with knowledge transfer from retinal vessels in fundus image. 
With the proposed vesselness-specific batch normalization and self-ensembling mean-teacher, it overcomes the challenging cross-anatomy domain shift and facilitate the annotation efficiency. Experimental results show that our method can effectively achieve accurate coronary artery segmentation with only very limited manual annotations required, bringing advantages for alleviating the annotation burden in clinical practice. In the further, it is of interest to leverage more different types of vessels for multi-source domain adaptation to further improve the segmentation accuracy and robustness. 

\section{COMPLIANCE WITH ETHICAL STANDARDS}
This study was performed in line with the principles of the Declaration of Helsinki. Approval was granted by the Ethics Committee of School of Biomedical Engineering in Shanghai Jiao Tong University (No. 2020004).

\section{Acknowledge}
This research is partially supported by Beijing Natural Science Foundation Haidian Original Innovation Collaborative Fund (No. L192006), 
the National Key research and development program (No.2016YFC0106200), the National Natural Science Foundation of China (No. 81771921 and No.61901084), and the funding from Institute of Medical Robotics of Shanghai Jiao Tong University as well as the 863 national research fund (No.2015AA043203). The authors declare that they have no conflict of interest.

\bibliographystyle{IEEEbib}
\bibliography{refs}

\end{document}